\documentclass[12pt]{article}
\usepackage{graphicx}

\def\Re{{\cal R \mskip-4mu \lower.1ex \hbox{\it e}\,}}
\def\Im{{\cal I \mskip-5mu \lower.1ex \hbox{\it m}\,}}
\def\ie{{\it i.e.}}

\def\etal{{\it et al.}}

\def\sub#1{_{\lower.25ex\hbox{$\scriptstyle#1$}}}
\def\tev{\,{\ifmmode\mathrm {TeV}\else TeV\fi}}
\def\gev{\,{\ifmmode\mathrm {GeV}\else GeV\fi}}
\def\mev{\,{\ifmmode\mathrm {MeV}\else MeV\fi}}
\def\mpl{\ifmmode \overline M_{Pl}\else $\overline M_{Pl}$\fi}
\def\to{\rightarrow}

\def\subw{_{\rm w}}
\def\mh{\ifmmode m\sbl H \else $m\sbl H$\fi}
\def\mch{\ifmmode m_{H^\pm} \else $m_{H^\pm}$\fi}
\def\mt{\ifmmode m_t\else $m_t$\fi}
\def\mc{\ifmmode m_c\else $m_c$\fi}
\def\mz{\ifmmode M_Z\else $M_Z$\fi}
\def\mw{\ifmmode M_W\else $M_W$\fi}
\def\mws{\ifmmode M_W^2 \else $M_W^2$\fi}
\def\mhs{\ifmmode m_H^2 \else $m_H^2$\fi}   
\def\mzs{\ifmmode M_Z^2 \else $M_Z^2$\fi}
\def\mts{\ifmmode m_t^2 \else $m_t^2$\fi}
\def\mcs{\ifmmode m_c^2 \else $m_c^2$\fi}
\def\mchs{\ifmmode m_{H^\pm}^2 \else $m_{H^\pm}^2$\fi}
\def\ztwo{\ifmmode Z_2\else $Z_2$\fi}
\def\zone{\ifmmode Z_1\else $Z_1$\fi}
\def\mtwo{\ifmmode M_2\else $M_2$\fi}
\def\mone{\ifmmode M_1\else $M_1$\fi}
\def\tb{\ifmmode \tan\beta \else $\tan\beta$\fi}
\def\xw{\ifmmode x\subw\else $x\subw$\fi}
\def\ch{\ifmmode H^\pm \else $H^\pm$\fi}
\def\lum{\ifmmode {\cal L}\else ${\cal L}$\fi}
\def\inpb{\,{\ifmmode {\mathrm {pb}}^{-1}\else ${\mathrm 
{pb}}^{-1}$\fi}}
\def\infb{\,{\ifmmode {\mathrm {fb}}^{-1}\else ${\mathrm 
{fb}}^{-1}$\fi}}
\def\epem{\ifmmode e^+e^-\else $e^+e^-$\fi}
\def\ppb{\ifmmode \bar pp\else $\bar pp$\fi}
\def\bsg{\ifmmode B\to X_s\gamma\else $B\to X_s\gamma$\fi}
\def\bsll{\ifmmode B\to X_s\ell^+\ell^-\else $B\to X_s\ell^+\ell^-$\fi}
\def\bstt{\ifmmode B\to X_s\tau^+\tau^-\else $B\to X_s\tau^+\tau^-$\fi}
\def\lamt{\ifmmode \tilde\lambda\else $\tilde\lambda$\fi}
\def\shat{\ifmmode \hat s\else $\hat s$\fi}
\def\that{\ifmmode \hat t\else $\hat t$\fi}
\def\uhat{\ifmmode \hat u\else $\hat u$\fi}

\newskip\zatskip \zatskip=0pt plus0pt minus0pt
\def\matth{\mathsurround=0pt}
\def\lsim{\mathrel{\mathpalette\atversim<}}
\def\gsim{\mathrel{\mathpalette\atversim>}}
\def\atversim#1#2{\lower0.7ex\vbox{\baselineskip\zatskip\lineskip\zatskip
  \lineskiplimit 
0pt\ialign{$\matth#1\hfil##\hfil$\crcr#2\crcr\sim\crcr}}}

\renewcommand{\thefootnote}{\fnsymbol{footnote}}

\begin{document} \begin{titlepage} 
\rightline{\vbox{\halign{&#\hfil\cr
&SLAC-PUB-9132\cr
}}}
\begin{center}

{\Large\bf Shifts in the Properties of the Higgs Boson from Radion 
Mixing}
\footnote{Work supported by the Department of 
Energy, Contract DE-AC03-76SF00515}
\medskip

\normalsize 
{\bf \large J.L. Hewett and T.G. Rizzo}
\vskip .3cm
Stanford Linear Accelerator Center \\
Stanford University \\
Stanford CA 94309, USA\\
\vskip .2cm

\end{center}

\begin{abstract} 
We examine how mixing between the Standard Model Higgs boson, $h$, and 
the radion present in the Randall-Sundrum model of localized gravity 
modifies the expected properties of the Higgs boson. In particular, we 
demonstrate that the total and partial decay widths of the Higgs, as 
well as the $h\to gg$ branching fraction, can be substantially altered 
from their Standard Model expectations.  The remaining branching 
fractions 
are modified less than $\lsim 5\%$ for most of the parameter space 
volume. 
\end{abstract} 

\renewcommand{\thefootnote}{\arabic{footnote}} \end{titlepage}


The Randall-Sundrum (RS) model of localized gravity {\cite {rs}} 
offers a potential solution to the 
hierarchy problem that can be tested at present and future 
accelerators {\cite {dhr}}.  In the original version of this model, 
the Standard Model (SM) fields are confined to one of two branes that 
are 
embedded in a 5-dimensional anti-de Sitter space (AdS$_5$)   
described by the metric $ds^2=e^{-2k|y|}\eta_{\mu\nu}dx^\mu 
dx^\nu-dy^2$,
with $y=r_c\phi$ where $r_c$ is the compactification radius and
$\phi$ describes the 5$^{th}$ dimension.
The parameter $k$ characterizes the curvature of the 5-dimensional 
space
and is naturally of order the Planck scale.  The two branes form the
boundaries of the AdS$_5$ slice and gravity is localized on the 
brane located at $y=0$.  Mass parameters on the Standard Model
brane, located at $y=r_c\pi$, are red-shifted compared to those on the
$y=0$ brane and are given 
by $\Lambda_\pi=\mpl e^{-kr_c\pi}$, where $\mpl$ is the reduced Planck
scale.  In order to address the hierarchy 
problem, $\Lambda_\pi\sim$ TeV and hence
the separation between the two branes, $r_c$, must have a value of 
$kr_c \sim 11-12$.   A number of authors \cite{gw} have demonstrated 
that
this quantity can be naturally stabilized by a mechanism which leads 
directly to the existence of a massive bulk scalar field.  

Fluctuations about the stabilized RS configuration allow for two 
massless
excitations described in the metric 
by $\eta_{\mu\nu}\to g_{\mu\nu}(x)$ and $r_c\to
T(x)$.  The first corresponds to the graviton and $T(x)$ is a new
scalar field arising from the $g_{55}$ component of the metric and is
known as the radion ($r_0$).  This scalar field corresponds 
to a quantum excitation of the separation between the two branes.  
The mass of the radion
is proportional to the backreaction of the bulk scalar vacuum
expectation value (vev) on the metric.
Generally, one expects that the radion mass should be in the range of  
a few~$\times 10$~GeV~$\leq m_{r_0} \leq \Lambda_\pi$, where the lower 
limit
arises from radiative corrections and the upper bound is the cutoff of
the effective field theory.  The radion mass $m_{r_0}$ is then expected 
to 
be below the scale $\Lambda_\pi$ implying that the radion may be 
the lightest new field present in the RS model. 
The radion couples to fields on the Standard Model brane via the trace 
of the 
stress-energy tensor with a strength $\Lambda \propto \Lambda_\pi$ of 
order the TeV scale, 
\begin{equation}
{\cal L}_{eff}=-r_0(x)~T^\mu_\mu /\Lambda\,.
\end{equation} 
Note that $\Lambda= \sqrt 3 \Lambda_\pi$ in the notation of 
Ref. {\cite {dhr}}.
This leads to gauge and matter couplings for the radion that 
are qualitatively similar to those of the SM Higgs boson. 
The collider production and decay of the RS radion has been examined 
by a number of authors {\cite {grw,big}} and has been recently reviewed 
by Kribs {\cite {Kribs}}. 

On general grounds of covariance, 
the radion may mix with the SM Higgs field, which is constrained to 
the TeV brane, through an interaction term of the form 
\begin{equation}
S_{rH}=-\xi \int d^4x \sqrt{-g_{ind}} R^{(4)}[g_{ind}] H^\dagger H\,.
\end{equation}
Here $H$ is the Higgs doublet field, 
$R^{(4)}[g_{ind}]$ is the 4-d Ricci scalar constructed out of the 
induced metric $g_{ind}$
on the SM brane,  and $\xi$ is a dimensionless mixing parameter assumed 
to be 
of order unity and with unknown sign.  The 
above action induces kinetic mixing between the $r_0$ and 
$h_0$ fields.  The resulting Lagrangian can be diagonalized by
a set of field redefinitions and rotations \cite{cgk},  
\begin{eqnarray}
h_0 & = & Ah+Br \,,\\
r_0 & = & Ch+Dr \,,\nonumber
\end{eqnarray}
with 
\begin{eqnarray}
A & = & \cos \theta -6\xi\gamma/Z \sin \theta \,,\\
B & = & \sin \theta +6\xi\gamma/Z \cos \theta \,, \nonumber \\
C & = & -\sin \theta/Z \,, \nonumber \\ 
D & = & \cos \theta/Z \,, \nonumber
\end{eqnarray}
where $h\,, r$ represent the physical fields, and
\begin{eqnarray}
\gamma & = & {v\over {\Lambda}}\,,\\
Z^2 & = & 1+6\xi(1-6\xi)\gamma^2\,,\nonumber
\end{eqnarray}
with $v\simeq 246$ GeV being the SM vacuum expectation value. The 
factor 
$Z$ serves to bring the physical radion kinetic term to canonical form 
and as 
such it must satisfy $Z>0$. For a fixed value of $\gamma$ this implies 
that 
the range of $\xi$ is bounded, \ie, $\xi_- \leq \xi \leq \xi_+$, where 
\begin{equation}
\xi_\pm= {1\over {12}}[1\pm(1+4/\gamma^2)^{1/2}]\,,
\end{equation}
For example, if $\gamma$ takes on the natural values
 $\gamma=0.2(0.1)$ then $\xi$ must lie in the approximate range 
$-0.754 \leq \xi \leq 0.921(-1.585 \leq \xi \leq 1.752)$. 
The masses of the physical states, $r,h$, are then given by
\begin{equation}
m_\pm^2={1\over {2}}\Big[T\pm\sqrt{T^2-4F}\Big]\,,
\end{equation}
where $m_+(m_-)$ is the larger(smaller) of the two masses and
\begin{eqnarray}
T & = & (1+t^2)m_{h_0}^2+m_{r_0}^2/Z\,,\\
F & = & m_{h_0}^2 m_{r_0}^2/Z^2\,,\nonumber
\end{eqnarray}
with $m_{h_0,r_0}$ being the weak interaction eigenstate masses 
and $t=6\xi\gamma/Z$. 
This mixing will clearly affect the phenomenology of both the radion 
and
Higgs fields.  In particular, the bounds on the Higgs mass from the 
standard
global fit to precision electroweak data are modified, allowing for a 
Higgs
boson (and radion) mass of order several hundred GeV \cite{cgk}.
Here, we examine the modifications to the properties of the Higgs 
boson,
in particular its decay widths and branching fractions, induced by this
mixing and find that substantial differences from the SM expectations 
can
be obtained.

\begin{figure}[htbp]
\centerline{
\includegraphics[width=9cm,angle=90]{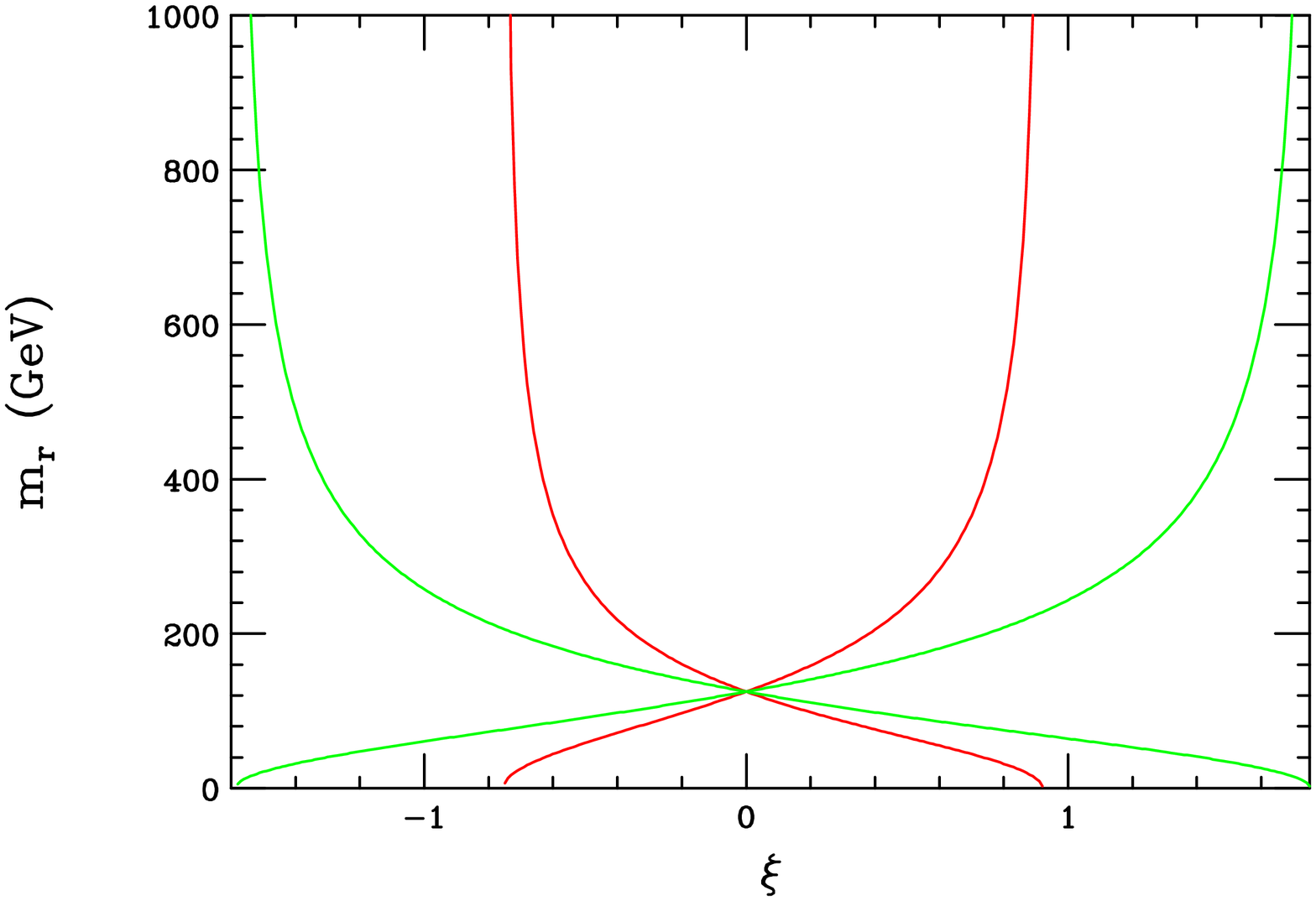}}
\vspace*{5mm}
\centerline{
\includegraphics[width=9cm,angle=90]{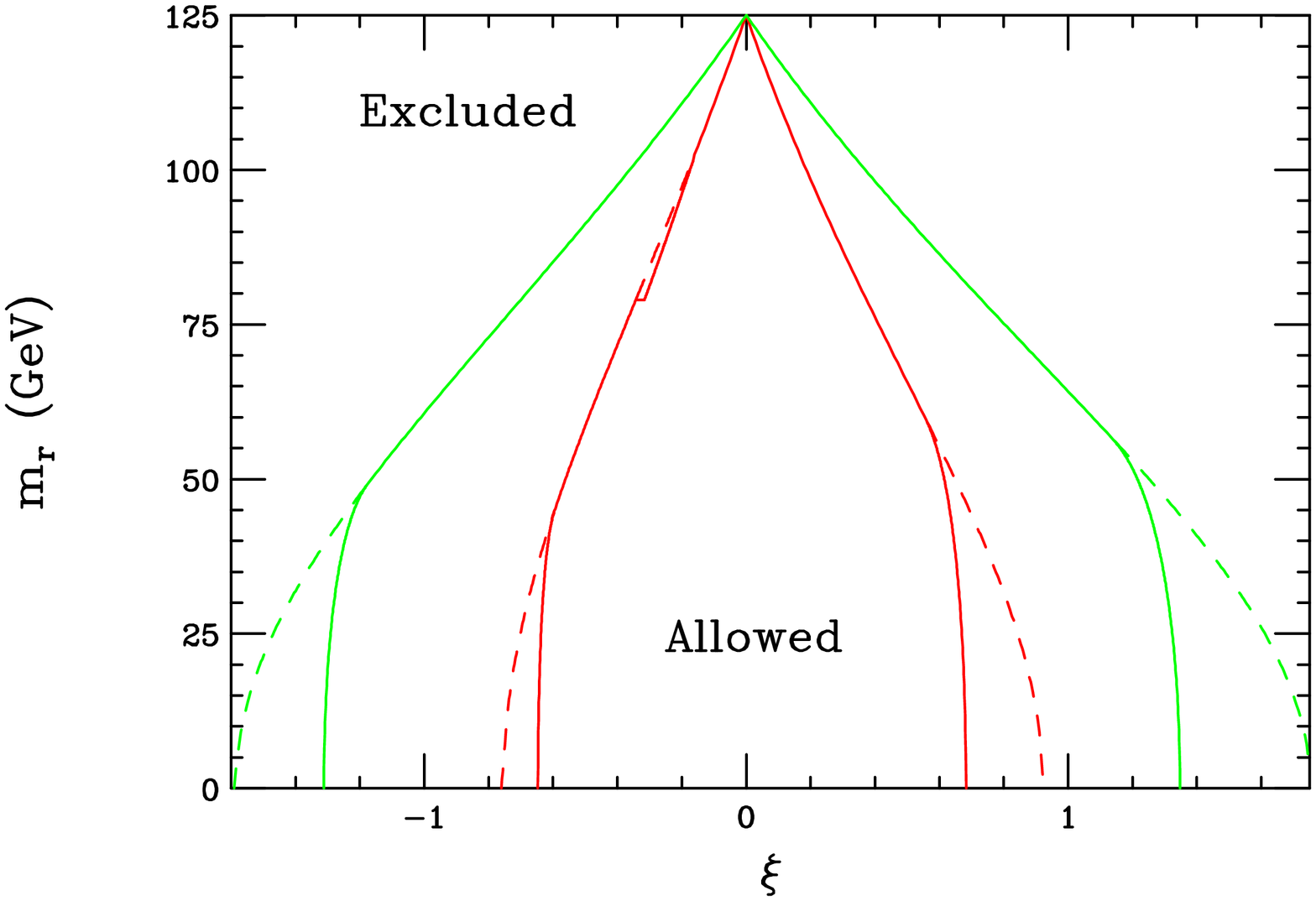}}
\vspace*{0.1cm}
\caption{Constraints on the mass of the radion assuming $m_h=125$ GeV 
as a 
function of $\xi$ as described in the text for 
$\gamma=0.1$(green, outer curves) and 
0.2(red, inner curves). The 
allowed region lies between the solid curves. In the lower 
panel the regions excluded by LEP 
searches are also shown and they lie between the corresponding 
solid and dashed curves.}
\label{fig1}
\end{figure}

To make predictions in this scenario we need to 
specify four parameters: the masses of the physical Higgs and radion 
fields, $m_{h,r}$,  the mixing 
parameter $\xi$, and the ratio $\gamma=v/\Lambda$.  Clearly this 
ratio cannot be too large as $\Lambda_\pi$ is already bounded 
from below by collider and electroweak precision 
data {\cite {dhr}}; Tevatron data for contact interactions 
yields the constraints $\Lambda_\pi\gsim
300,\, 1500,\, 4500$ GeV for $k/\mpl=1.0,\, 0.1,\, 0.01$, respectively.
Curvature constraints suggest that $k/\mpl\leq 0.1$ \cite{dhr}, which 
in
turn implies $v/\Lambda\lsim 0.1$.
For definiteness we will relax this bound somewhat and  take 
$v/\Lambda \leq 0.2$ along with a physical Higgs mass of 125 GeV 
in our analysis.
We note that large absolute values of $\xi$ 
and wide ranges of $v/\Lambda$ have been entertained in the literature.  

The values of the two physical masses themselves are not arbitrary. 
When we require the weak eigenstate 
mass-squared parameters of the unmixed radion and Higgs 
fields to be real, as is demanded by hermiticity, we obtain an 
additional 
constraint on the ratio of the 
physical radion and Higgs masses which depends on both   
$\xi$ and $\gamma$. Defining the ratio $r=(m_+/m_-)^2$, one finds that 
these 
two conditions require that $r$ must be bounded from below by 
\begin{equation}
r_{min}=1+2t^2\pm 2|t|\sqrt{1+t^2}\,,
\end{equation}
where the sign depends on the ordering of $m_h$ and $m_r$.
Note that this bound implies that 
it is disfavored for the radion to have a mass near 
that of the Higgs when there is significant mixing.
The resulting excluded region is shown in Fig. \ref{fig1} for the
demonstrative case of $m_h=125$ GeV.   These
constraints are found to be somewhat restrictive.
The low mass region for the radion is also 
partially excluded by direct searches
at LEP as can be seen from the figure.  In calculating the LEP 
constraints, we fixed $v/\Lambda$ to the conservative values of
$v/\Lambda=0.2(0.1)$ while varying $\xi$,
and converted the LEP Higgs search bounds {\cite {sop}} into 
constraints 
for the radion using the appropriate set of rescaling factors. While 
some of the low mass region for the radion is eliminated for these 
values of the parameters, the 
parameter space for a light radion is certainly not closed. As one 
decreases the assumed fixed value of $v/\Lambda$, the size of the 
allowed 
low mass region 
grows since the radion couplings to the $Z$ boson are rapidly 
shrinking. 

Once $\xi, \gamma $ and $m_{h,r}$ are specified, the angle $\theta$ 
describing the rotations into the Higgs-radion physical states becomes 
calculable; one finds that
\begin{eqnarray}
\tan 2\theta & = & {2tZ^2m_{h_0}^2\over 
{m_{r_0}^2-m_{h_0}^2(Z^2-36\xi^2
\gamma^2)}}\,, \\
\sin 2\theta & = & {2tm_{h_0}^2\over {m_r^2-m_h^2}}\,. \nonumber
\end{eqnarray}
The weak basis masses are given by
\begin{equation}
m_{h_0}^2={{m_+^2+m_-^2\pm 
\sqrt{(m_+^2+m_-^2)^2-(2m_+m_-)^2(1+t^2)}}\over 
{2(1+t^2)}}\,,
\end{equation}
with the sign chosen positive(negative) when $m_h(m_r)$ is identified 
with 
$m_\pm$.  In either case one obtains $m_{r_0}=Zm_+ m_-/m_{h_0}$.   
\begin{figure}[htbp]
\centerline{
\includegraphics[width=9cm,angle=90]{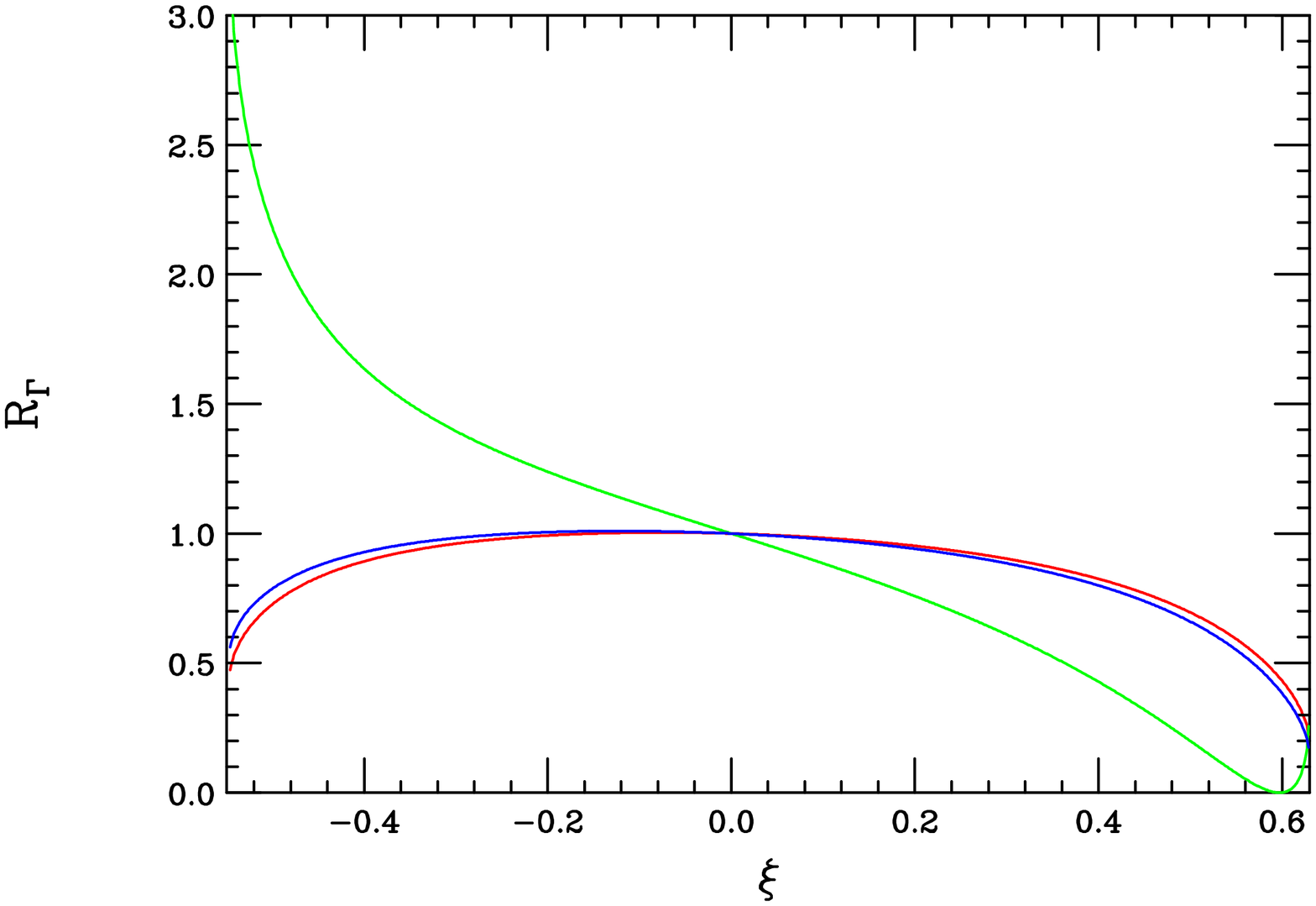}}
\vspace*{5mm}
\centerline{
\includegraphics[width=9cm,angle=90]{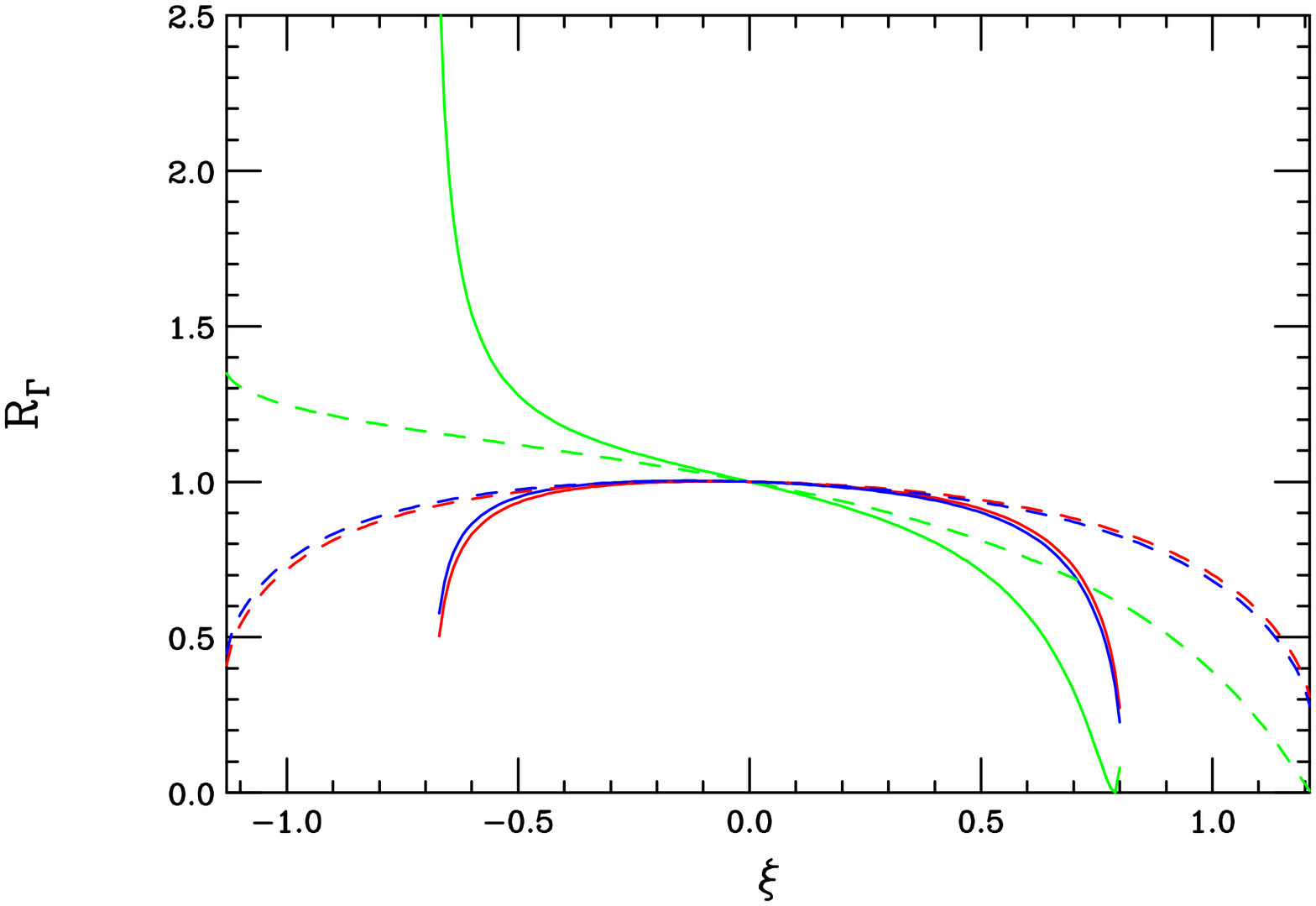}}
\vspace*{0.1cm}
\caption{Ratio of Higgs partial 
widths to their SM values, $R_\Gamma$, as a function 
of $\xi$ assuming a physical Higgs mass of 125 GeV: red for fermion 
pairs or 
massive gauge boson pairs, green for gluons and blue for photons.  This
corresponds to gluons, photons, $V\bar V$ from bottom to top on the 
right. 
In the top  
panel we assume $m_r=300$ GeV and $v/\Lambda=0.2$. In the bottom panel 
the 
solid(dashed) curves are for $m_r=500(300)$ GeV and 
$v/\Lambda=0.2(0.1)$.}
\label{fig2}
\end{figure}

We now turn our attention to the properties of the Higgs boson in this
model when mixing with the radion is included. 
Following the notation of Csaki \etal~ {\cite {cgk}}, the couplings of 
the 
physical Higgs to the SM fermions and massive gauge bosons $V=W,Z$ 
is now given by
\begin{equation}
{\cal {L}}={-1\over {v}}(m_f\bar ff-m_V^2 V_\mu V^\mu)[\cos \theta-t
\sin \theta-{v\over {\Lambda}}\sin \theta/Z]h\,,
\end{equation}
where the angle $\theta$ is defined above and 
can now be calculated in terms of the four 
parameters $\xi$, $v/\Lambda$, and the physical Higgs and radion 
masses. 
Note that the shifts to the fermionic and massive gauge boson
couplings are identical.
Denoting the combinations $\alpha=\cos \theta-t\sin \theta$ and 
$\beta=-\sin \theta/Z$, the corresponding Higgs 
coupling to gluons can be written as 
\begin{equation}
{\cal L} = c_g {\alpha_s\over {8\pi}}G_{\mu\nu}G^{\mu\nu}h\,,
\end{equation}
with
\begin{equation} 
c_g={-1\over {2v}}[(\alpha +{v\over {\Lambda}}\beta)F_g
-2b_3\beta {v\over {\Lambda}}]\,.
\end{equation}
Here, the first term is the usual one-loop top-quark contribution to
the $ggh$ coupling, whereas the second term arises from the trace 
anomaly
and appears solely from the mixing.  $b_3=7$ is the $SU(3)$ 
$\beta$-function 
and $F_g$ is a well-known kinematic function of the ratio of masses of 
the  
top-quark to the physical Higgs boson \cite{hhg}. 
Similarly the physical Higgs coupling to two photons is now given by 
\begin{equation}
{\cal L} = c_\gamma {\alpha_{em}\over {8\pi}}F_{\mu\nu}F^{\mu\nu}h\,,
\end{equation}
where
\begin{equation} 
c_\gamma={1\over {v}}[-(\alpha +{v\over {\Lambda}}\beta)F_\gamma +
(b_2+b_Y)\beta {v\over {\Lambda}}]\,.
\end{equation}
Here, $b_2=19/6$ and $b_Y=-41/6$ are the 
$SU(2)\times U(1)$ $\beta$-functions and $F_\gamma$ is another 
well-known 
kinematic function \cite{hhg} of the ratios of the $W$ boson and 
top-quark 
masses to the physical 
Higgs mass, and second term again originates from the trace anomaly.
Note that in the simultaneous limits $\alpha \to 1,~\beta \to 0$ 
we recover the usual SM results. From these expressions we can now 
compute  
the modifications to the various decay widths and branching fractions 
of the
SM Higgs due to mixing with the radion.  

Fig.~\ref{fig2} shows the ratio of the various Higgs partial widths in 
comparison to their SM expectations as a function of the parameter 
$\xi$ 
for sample values of $m_r$ and ${v\over {\Lambda}}$
and assuming that $m_h=125$ GeV.  The range of the curves reflects the 
allowed parameter region for $\xi$.
Several features are immediately apparent: ($i$) the shifts in 
the widths to $\bar ff/VV$ and $\gamma \gamma$ final states are very 
similar; 
this is due to the relatively large magnitude of $F_\gamma$ while the 
combination $b_2+b_Y$ is rather small. ($ii$) On the otherhand, the 
shift for 
the $gg$ final state can be substantial; here, $F_g$ is numerically 
smaller than $F_\gamma$ and $b_3$ is quite large, resulting in a large 
contribution from the trace
anomaly term. ($iii$) For relatively light radions with a lower 
value of $\Lambda$, the Higgs decay 
width into the $gg$ final state can be vanishingly small.  This is 
due to a strong 
destructive interference between the two contributions to the 
amplitude for values of $\xi$ near $-1$. ($iv$) Increasing the value of 
$m_r$ 
has less of an effect than does a decrease in the ratio 
${v\over {\Lambda}}$. 

\begin{figure}[htbp]
\centerline{
\includegraphics[width=5.4cm,angle=90]{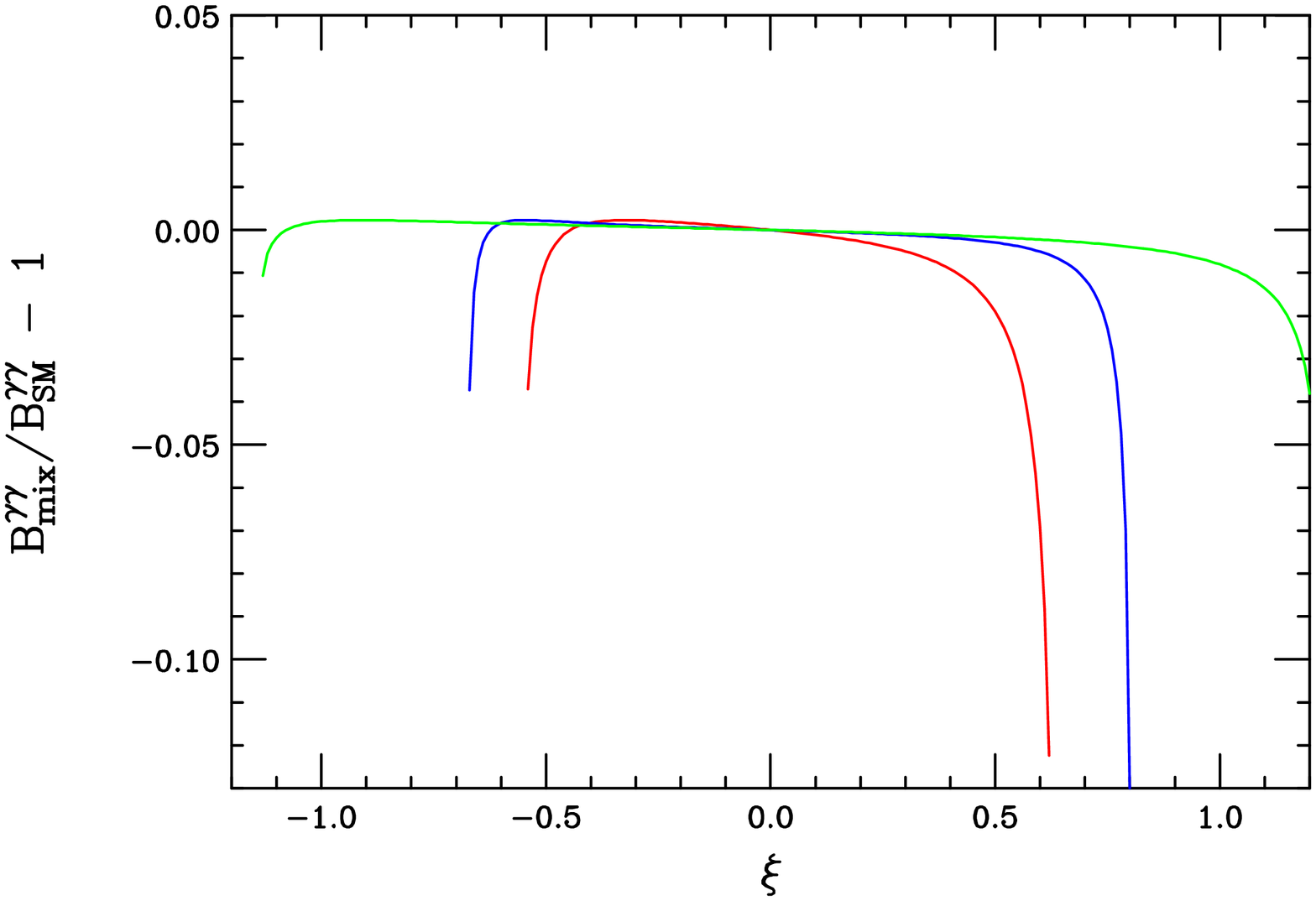}
\hspace*{5mm}
\includegraphics[width=5.4cm,angle=90]{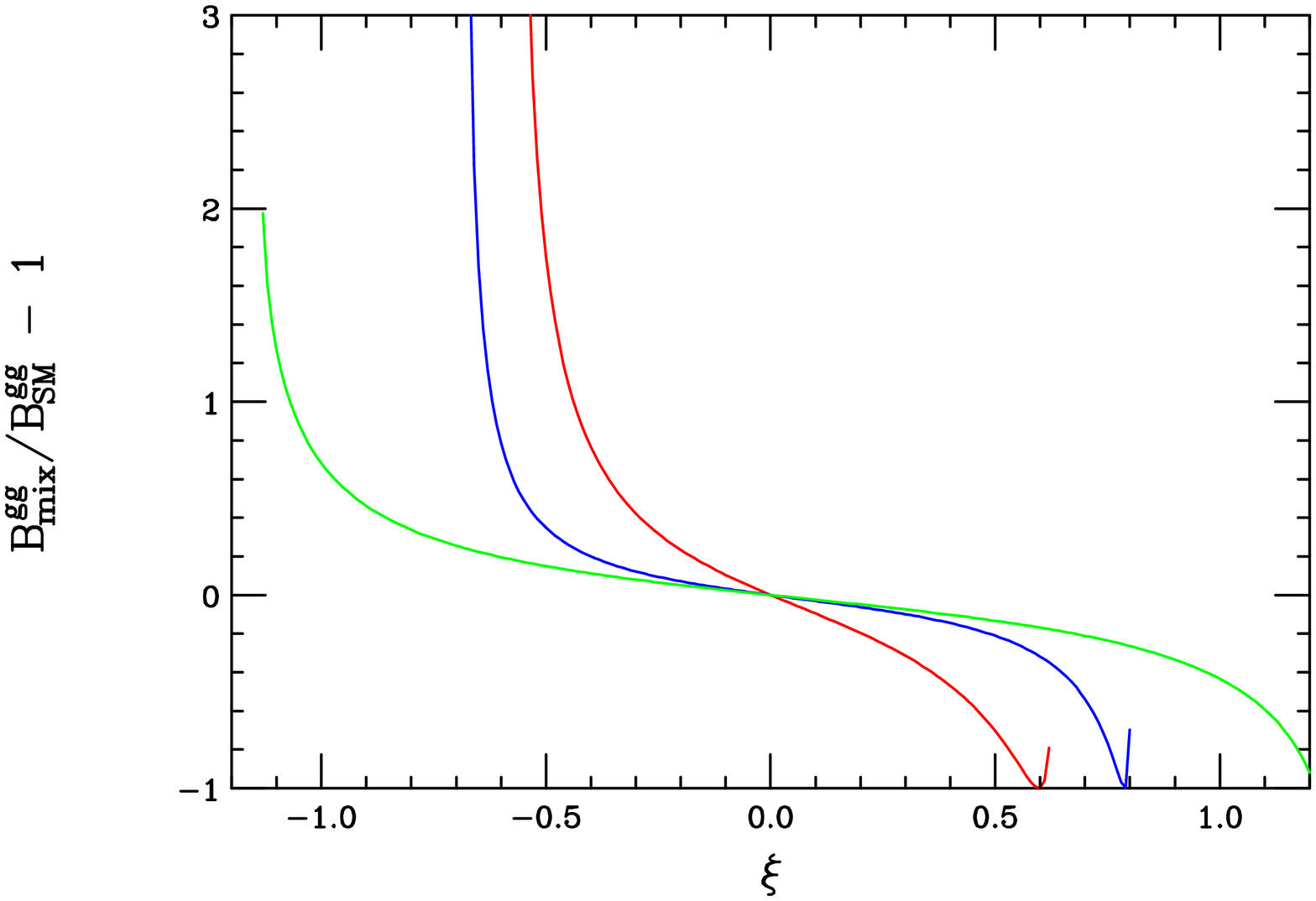}}
\vspace*{0.1cm}
\centerline{
\includegraphics[width=5.4cm,angle=90]{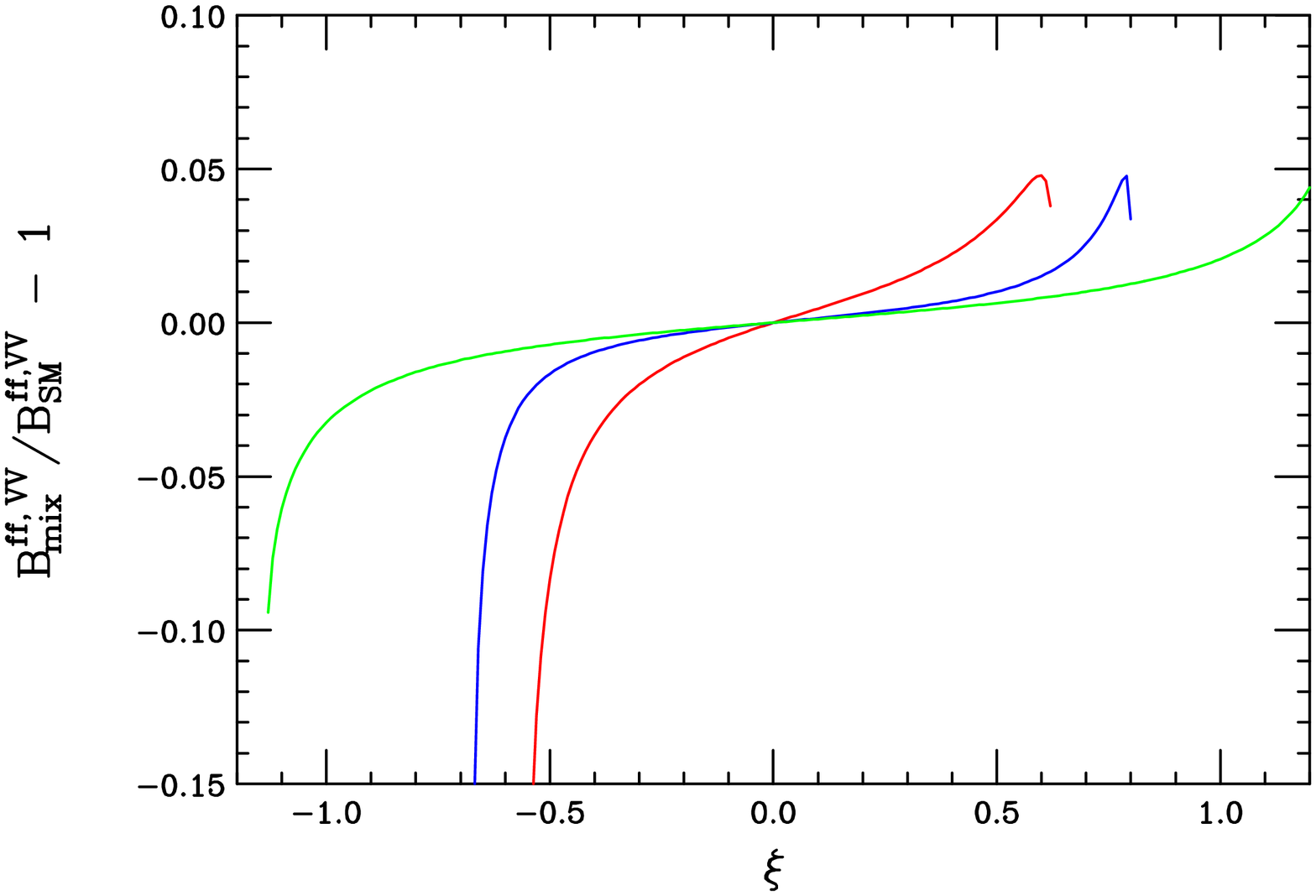}
\hspace*{5mm}
\includegraphics[width=5.4cm,angle=90]{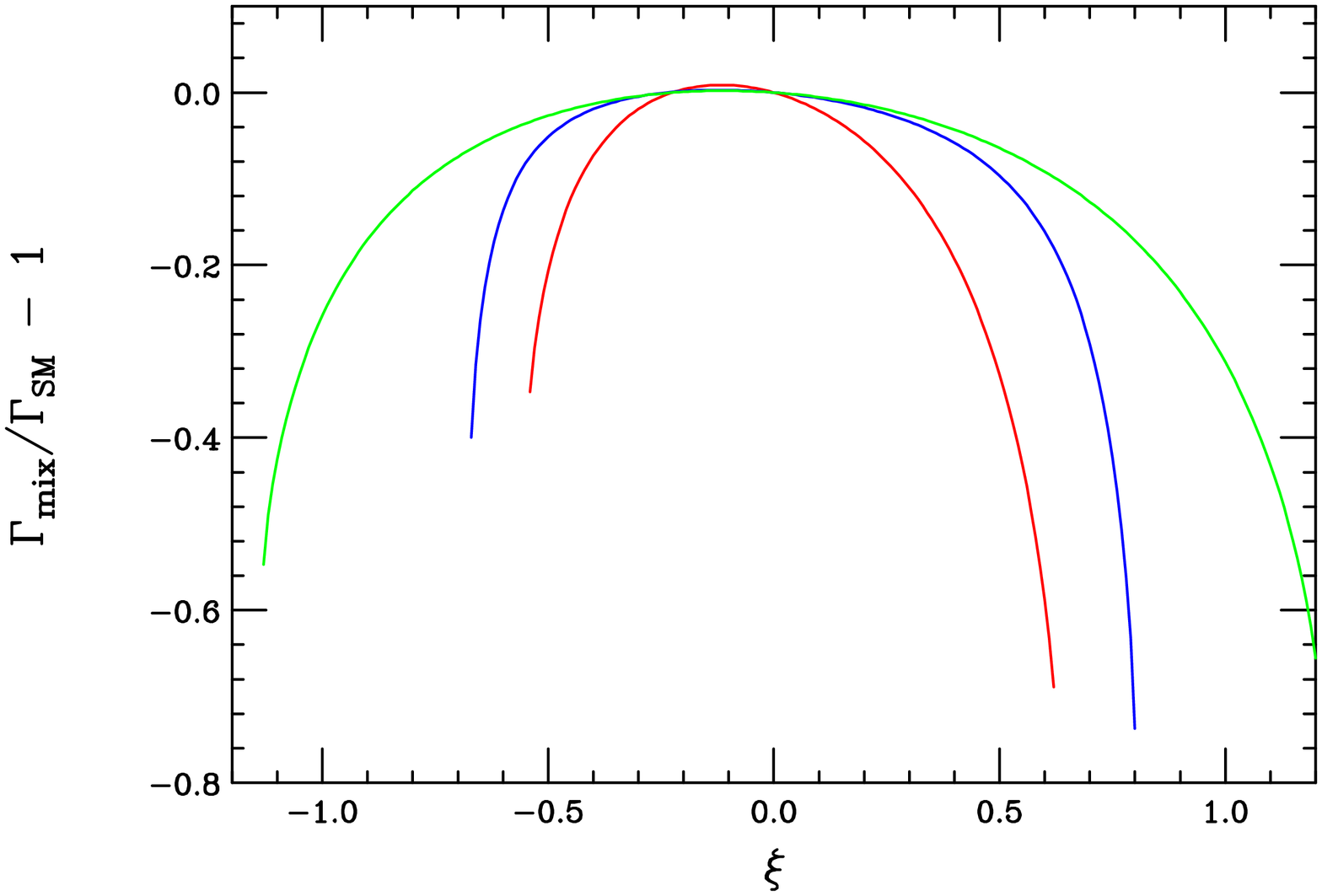}}
\vspace*{0.1cm}
\caption{The ratio of the Higgs
branching fraction into $\gamma\gamma$, $gg$, $f\bar f$, and $VV$ final
states as labeled, as well as for the total width, with radion mixing
to that of the SM as a function of $\xi$.  The red (blue; green) 
curves represent the choice $m_r=300$ GeV, $v/\Lambda=0.2$
(500, 0.2; 300, 0.1), and correspond to the curves which have the most
limited (central; largest) range of $\xi$, respectively, in the figure.}
\label{fig3}
\end{figure}

\begin{figure}[htbp]
\centerline{
\includegraphics[width=9cm,angle=90]{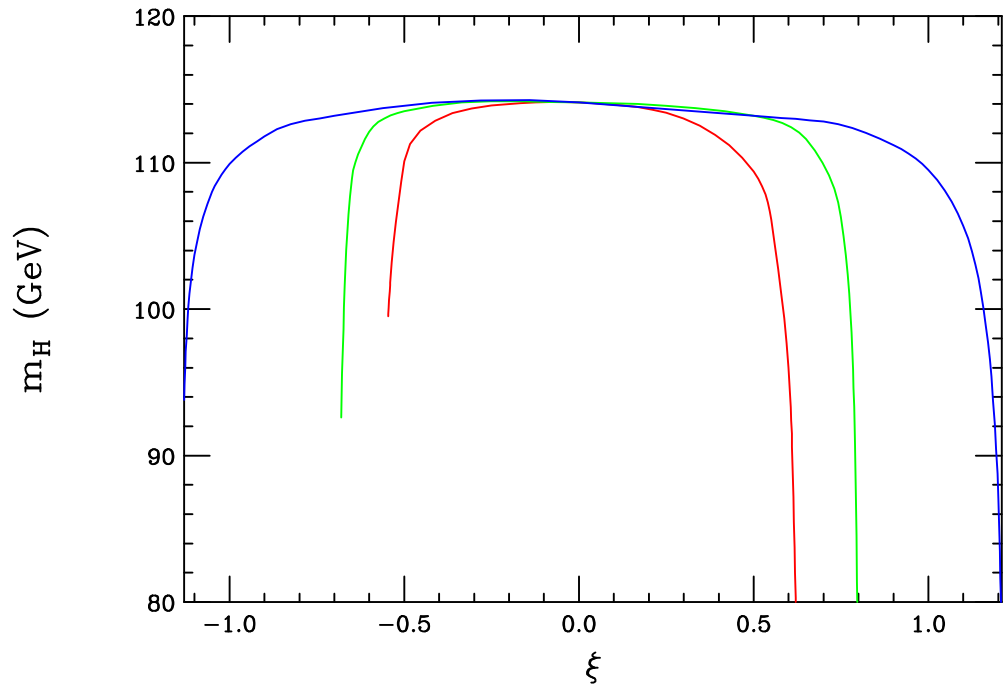}}
\vspace*{0.1cm}
\caption{Lower bound on the mass of the Higgs boson from direct 
searches at 
LEP as a function of $\xi$ including the effects of mixing. The red 
(green; blue) 
curves correspond to the choice $m_r=300$ GeV, $v/\Lambda=0.2$
(300, 0.1; 500, 0.2) and represent the curves from bottom to top.}
\label{fig4}
\end{figure}

The  deviation from SM expectations for
the various branching fractions, as well as the total width, of the 
Higgs are displayed in Fig. \ref{fig3} as a function of the 
mixing parameter $\xi$.  As above, the range of the curves reflects the
allowed parameter region for $\xi$.
  We see that the gluon branching fraction and the
total width may be drastically different than that of the SM.  As we 
will 
see below, the former may greatly affect the Higgs production cross 
section 
at the LHC.  However, the
$\gamma\gamma$, $f\bar f$, and $VV$ (where $V=W,Z$) branching fractions
only receive small corrections to their SM values, of order $\lsim 
5-10\%$
for almost all of the parameter region except near the edges of the 
parameter 
space.  Observation of these shifts 
will require the precise determination of the Higgs branching fractions 
which is obtainable at a future high energy $e^+e^-$ Linear 
Collider {\cite {bat}}.  Once these measurements are performed, 
constraints on the radion parameter 
space may be extracted as will be discussed below. 
These small changes in the $ZZh$ and $hb\bar b$ 
couplings of the Higgs boson can also lead to small reductions in the 
Higgs 
search reach from LEPII. This is presented in Fig.~\ref{fig4} for 
several sets of parameters; except for extreme parameter cases this 
reduction 
in reach is rather modest. 

\begin{figure}[htbp]
\centerline{
\includegraphics[width=9cm,angle=90]{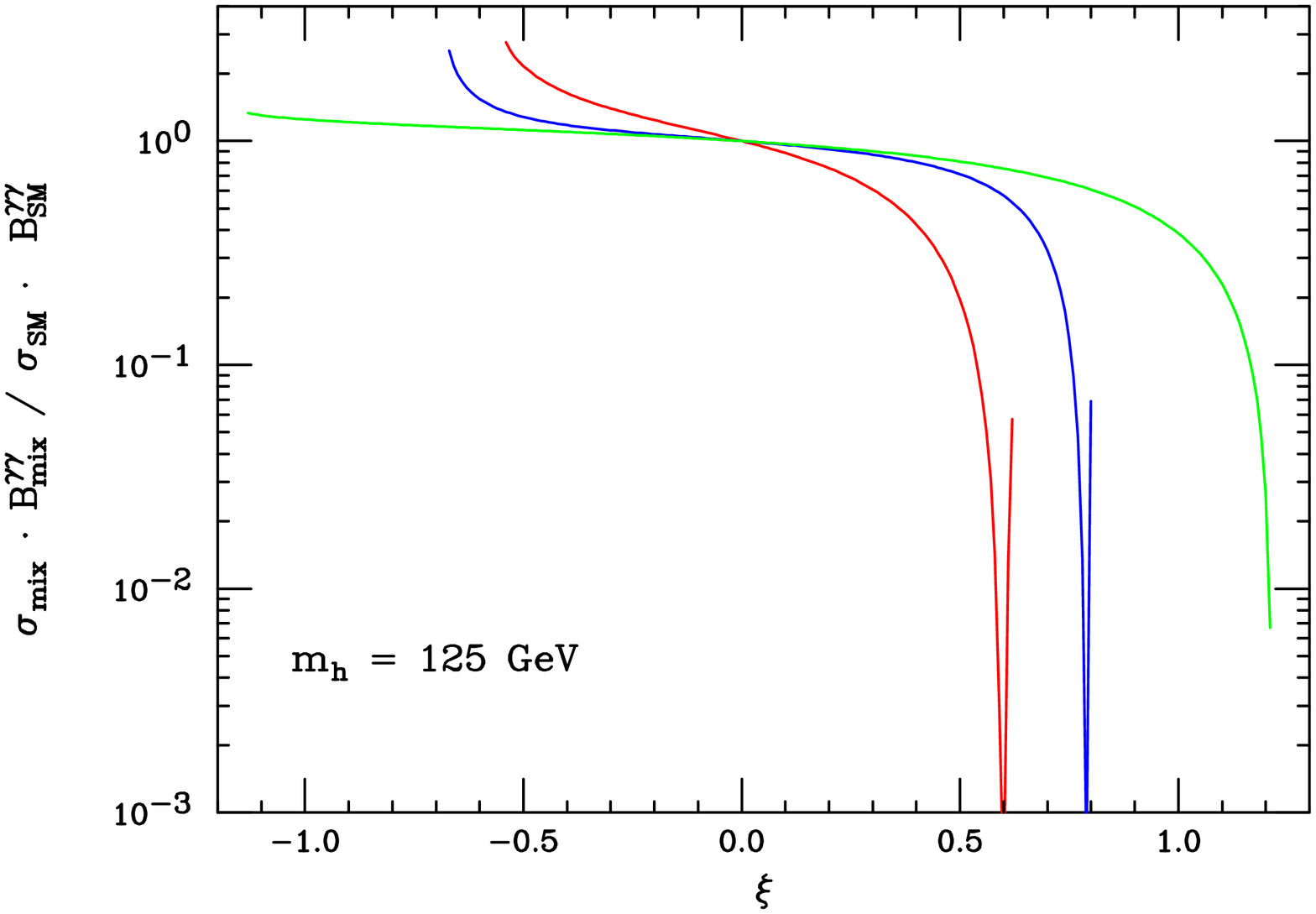}}
\vspace*{0.1cm}
\caption{The ratio of production cross section times branching fraction
for $pp\to h\to\gamma\gamma$ via gluon fusion with radion mixing to the
SM expectations as a function of $\xi$.  The Higgs mass is taken to be 
125 GeV.  The red (blue; green) 
curves correspond to the choice $m_r=300$ GeV, $v/\Lambda=0.2$
(500, 0.2; 300, 0.1), from left to right on the RHS of the figure.}
\label{fig5}
\end{figure}

At the LHC, the dominant production mechanism and signal for a
light Higgs boson 
is gluon-gluon fusion through a triangle graph with subsequent decay 
into $\gamma \gamma$. Both the production cross section and the  
$\gamma \gamma$ branching fraction are modified by mixing with the 
radion, leading to the results illustrated in Fig.~\ref{fig5}. 
This figure shows that the 
Higgs production rate in this channel
at the LHC can be either significantly reduced or somewhat enhanced 
in comparison to the expectations of the SM due 
to the effects of mixing. For some values of the parameters the 
reduction can 
be by more than an order of magnitude which could seriously hinder the
discovery of the Higgs via this channel at the LHC.

\begin{figure}[htbp]
\centerline{
\includegraphics[width=9cm,angle=90]{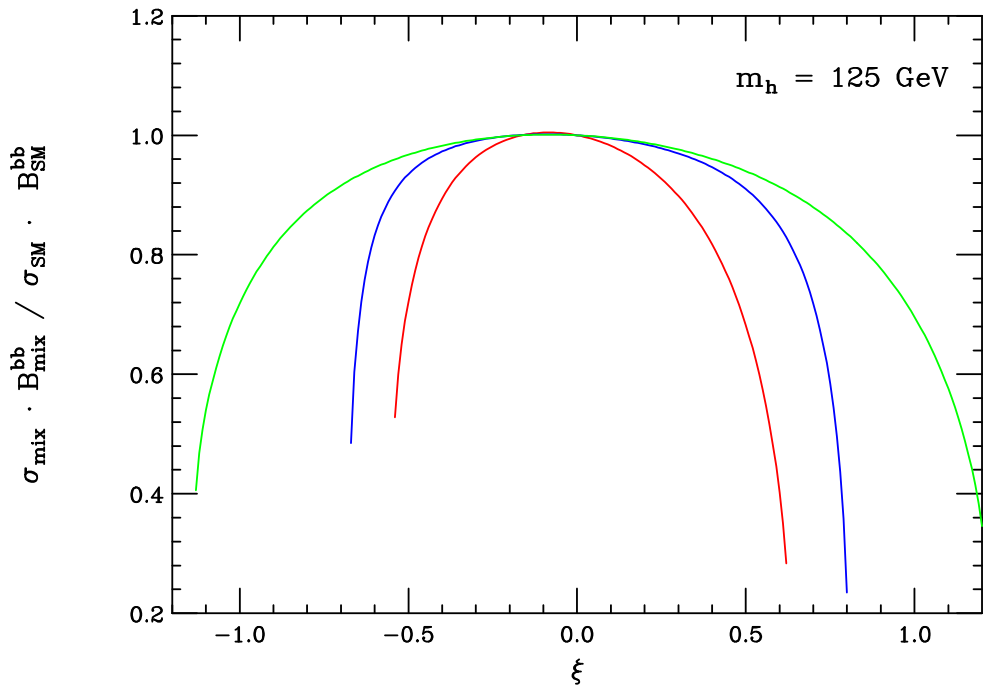}}
\vspace*{0.1cm}
\caption{The ratio of production cross section times branching fraction
for $\gamma\gamma\to h\to b\bar b$ with radion mixing to the
SM expectations as a function of $\xi$.  The Higgs mass is taken to be 
125 GeV.  The red (blue; green) 
curves, corresponding to the bottom (middle, top) curves, represent
the parameter choices $m_r=300$ GeV, $v/\Lambda=0.2$
(500, 0.2; 300, 0.1).}
\label{fig6}
\end{figure}

The s-channel production of the Higgs boson at a high energy 
photon photon collider
is an important channel for determining the properties of the Higgs.
In Fig. \ref{fig6} we see that decreases in
the production rate of the Higgs compared to SM expectations 
can also occur in the reaction
$\gamma\gamma\to h\to b\bar b$ when mixing with the radion is
included.  Again, once such mixing is taken into account, 
the event rate is reduced, possibly hampering 
the statistical ability of a future high energy photon collider to
measure the Higgs properties.  However, note that the potential
reductions in rate for this channel are not as drastic as those
which may be realized at the LHC.

\begin{figure}[htbp]
\centerline{
\includegraphics[width=9cm,angle=90]{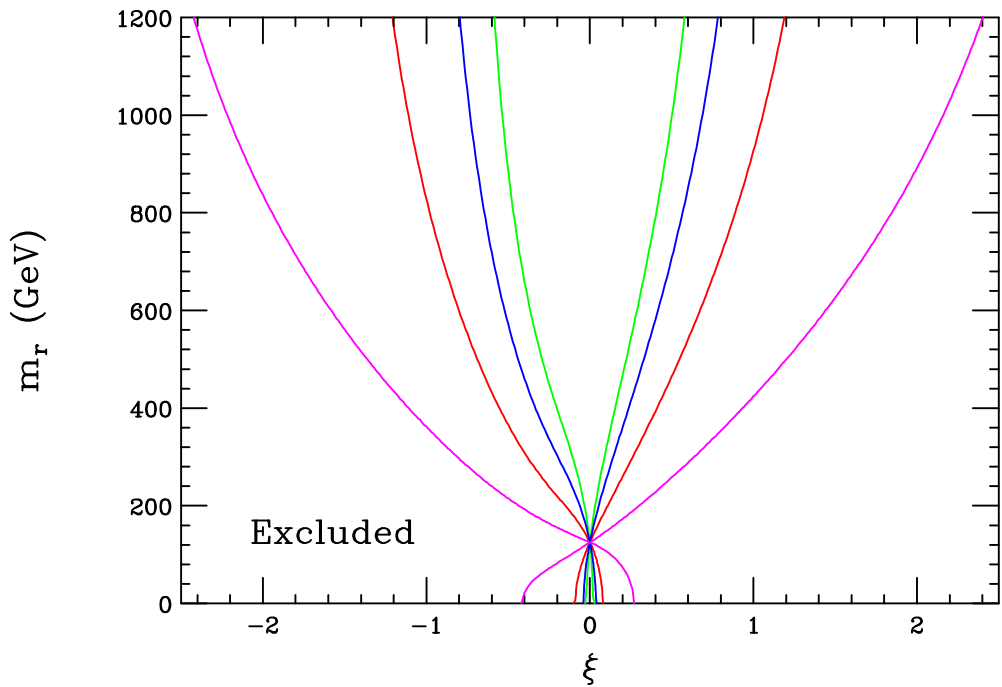}}
\vspace*{0.1cm}
\caption{95\% CL indirect bounds on the mass and mixing of the radion
arising from the precision measurements of the Higgs couplings 
obtainable
at the LHC and a Linear Collider for a Higgs mass of 125 GeV.  The
allowed region lies between the corresponding pair of vertical curves.
From the inner to outer curves, they represent the parameter values
$v/\Lambda=0.2,\, 0.15,\, 0.10,$ and $0.05$, respectively.}
\label{fig7}
\end{figure}

Once data from both the LHC and the Linear Collider (LC) is available, 
the 
radion parameter space can be explored using both direct searches and 
indirect
measurements.  In fact, the precision measurements of the Higgs boson 
couplings discussed above can be used to constrain the radion
parameter space beyond what may be possible via direct searches for
the radion.  For purposes of demonstration, let us assume that the LHC
and LC determine, within their capabilities, that the Higgs couplings 
are consistent with the predictions of the SM.  Using the expectations
for Higgs production at the LHC and LC from the analyses contained in
Ref. \cite{bat}, we derive the resulting excluded regions in the $\xi - 
m_r$
parameter plane.  This is displayed in Fig. \ref{fig7} for various
values of $v/\Lambda$ for the sample case of $m_h=125$ GeV.  In this
figure, the allowed region lies between the corresponding pair of
vertical curves.  Here, we see that if such a set of measurements of
the Higgs properties are realized, then a large fraction of the
radion parameter space would be excluded.  Direct radion searches
at these colliders would fully cover the lower portion of the remaining
parameter space up to the radion search reach.  Together, the direct
and indirect constraints would then only allow for a high mass radion 
with 
small mixing as a possibility under this scenario. 

Lastly, we note that the patterns of the shifts in the Higgs boson
properties due to radion mixing are distinct from those in other
models of new physics.  For example, consider the two Higgs
double model \cite{hhg}.  In this model, the ratio of the $hVV$
coupling to its SM value is given by $\sin(\beta-\alpha)$ and the 
similar
ratio for the $ht\bar t$ coupling is $\cos\alpha/\sin\beta$, where 
$\tan\beta$ is the ratio of vevs of the two Higgs doublets and $\alpha$
is the mixing angle between the two doublets.  In one variant of this
model, the corresponding ratio of the $hb\bar b$ coupling is equal, up
to a sign, of that for the $ht\bar t$ coupling, while in a second 
version
of the model the $hb\bar b$ ratio is $-\sin\alpha/\cos\beta$.  In 
either case, this spectrum of couplings cannot be reproduced via
radion mixing.

In summary, we see that Higgs-radion mixing, which is present in some
extra dimensional scenarios, can have a substantial effect on the
properties of the Higgs boson.  These modifications affect the total 
and
partial widths, as well as the
branching fractions, of Higgs decay into various final states.  This,
in turn, can significantly alter the expectations for Higgs production 
at LEP, the LHC, and a photon collider. 
For some regions of the parameters, the size of these shifts in the 
Higgs
widths and branching 
fractions may require the precision of a Linear Collider in order to be
studied in detail.

\bigskip

Note Added: A related paper \cite{Dominici:2002jv}
appeared on the arXiv four months after
this work and one year after a preliminary version of this work was
presented at summer conferences \cite{Hewett:2002rz}.

%
\def\MPL #1 #2 #3 {Mod. Phys. Lett. {\bf#1},\ #2 (#3)}
\def\NPB #1 #2 #3 {Nucl. Phys. {\bf#1},\ #2 (#3)}
\def\PLB #1 #2 #3 {Phys. Lett. {\bf#1},\ #2 (#3)}
\def\PR #1 #2 #3 {Phys. Rep. {\bf#1},\ #2 (#3)}
\def\PRD #1 #2 #3 {Phys. Rev. {\bf#1},\ #2 (#3)}
\def\PRL #1 #2 #3 {Phys. Rev. Lett. {\bf#1},\ #2 (#3)}
\def\RMP #1 #2 #3 {Rev. Mod. Phys. {\bf#1},\ #2 (#3)}
\def\NIM #1 #2 #3 {Nuc. Inst. Meth. {\bf#1},\ #2 (#3)}
\def\ZPC #1 #2 #3 {Z. Phys. {\bf#1},\ #2 (#3)}
\def\EJPC #1 #2 #3 {E. Phys. J. {\bf#1},\ #2 (#3)}
\def\IJMP #1 #2 #3 {Int. J. Mod. Phys. {\bf#1},\ #2 (#3)}
\def\JHEP #1 #2 #3 {J. High En. Phys. {\bf#1},\ #2 (#3)}

\end{document}